\documentstyle[12pt,twoside]{article}
\def\greaterthansquiggle{\raise.3ex\hbox{$>$\kern-.75em\lower1ex\hbox{$\sim$}}}

\def\lessthansquiggle{\raise.3ex\hbox{$<$\kern-.75em\lower1ex\hbox{$\sim$}}}

\newcommand{\beq}{\begin{equation}}
\newcommand{\eeq}{\end{equation}}
\newcommand{\beqa}{\begin{eqnarray}}
\newcommand{\eeqa}{\end{eqnarray}}
\newcommand{\beqan}{\begin{eqnarray*}}
\newcommand{\eeqan}{\end{eqnarray*}}
\newcommand{\ba}{\begin{array}}
\newcommand{\ea}{\end{array}}

\newcommand{\ra}{\rightarrow}

\newcommand{\wt}{\widetilde}
\newcommand{\wh}{\widehat}

\newcommand{\A}{{\cal A}}

\newcommand{\D}{{\cal D}}

\newcommand{\F}{{\cal F}}
\newcommand{\G}{{\cal G}}
\newcommand{\Ha}{{\cal H}}

\newcommand{\M}{{\cal M}}

\newcommand{\U}{{\cal U}}

\newcommand{\st}{\stackrel}
\newcommand{\dfrac}{\displaystyle \frac}

\def\nz{\ifmmode {I\hskip -3pt N} \else {\hbox {$I\hskip -3pt N$}}\fi}
\def\zz{\ifmmode {Z\hskip -4.8pt Z} \else
       {\hbox {$Z\hskip -4.8pt Z$}}\fi}
\def\qz{\ifmmode {Q\hskip -5.0pt\vrule height6.0pt depth 0pt
       \hskip 6pt} \else {\hbox
       {$Q\hskip -5.0pt\vrule height6.0pt depth 0pt\hskip 6pt$}}\fi}
\def\rz{\ifmmode {I\hskip -3pt R} \else {\hbox {$I\hskip -3pt R$}}\fi}
\def\cz{\ifmmode {C\hskip -4.8pt\vrule height5.8pt\hskip 6.3pt} \else
       {\hbox {$C\hskip -4.8pt\vrule height5.8pt\hskip 6.3pt$}}\fi}
\def\au{{\setbox0=\hbox{\lower1.36775ex%
\hbox{''}\kern-.05em}\dp0=.36775ex\hskip0pt\box0}}
\def\ao{{}\kern-.10em\hbox{``}}
\def\lint{\int\limits}
\renewcommand{\baselinestretch}{1.5} % Zeilenabstand
\normalsize
\begin{document}
\bibliographystyle{plain}

\begin{titlepage}
\begin{flushright}
UWThPh-1999-52 \\
\end{flushright}
\vspace{2cm}
\begin{center}
{\Large \bf Global Aspects of Quantizing 
\\[5pt]
Yang-Mills Theory} \\[40pt]
$ \mbox{Helmuth H\"uffel}^{1,2} \mbox{\, and Gerald Kelnhofer}^{3}$\\
Institut f\"ur Theoretische Physik \\
Universit\"at Wien \\
Boltzmanngasse 5, A-1090 Vienna, Austria
\vfill

{\bf Abstract}
\end{center}
\renewcommand{\baselinestretch}{1.0} % Zeilenabstand
\small

We review recent results on the derivation of a global path integral 
density for 
Yang-Mills theory. Based on a generalization of the stochastic 
quantization scheme and its geometrical interpretation we first  
 recall how locally
a modified Faddeev-Popov path integral density for the 
quantization of Yang-Mills 
theory can be derived, the modification consisting in the presence of 
specific 
$finite$ contributions of the pure gauge degrees of freedom. Due to the 
Gribov 
problem the gauge fixing can be defined only locally and the whole space 
of 
gauge potentials has to be partitioned into patches. We discuss a global 
extension of the path 
integral by summing over all patches, which can be proven to 
be
manifestly independent of the specific local choices of patches and
gauge 
fixing 
conditions, respectively. In addition to the formulation on the whole 
space of gauge potentials we discuss the corresponding path integral on 
the gauge orbit space.

\vfill
\begin{enumerate}
\item[1)] Talk given by Helmuth H\"uffel at the Central European
Triangle 
Meeting on Particle Physics, Zagreb, Croatia, June 17-19, 1999
\item[2)] Email: helmuth.hueffel@univie.ac.at
\item[3)] Supported by "Fonds zur F\"orderung der wissenschaftlichen 
Forschung in \"Oster\-reich",  project P10509-NAW
\end{enumerate}
\end{titlepage}
\renewcommand{\baselinestretch}{1.5} % Zeilenabstand
\normalsize

%*******************************************************************
The Faddeev-Popov \cite{Fadd} path integral procedure constitutes one of 
the most popular quantization methods for Yang-Mills theory and is
widely 
used in elementary particle physics. It is, however,  well known that at
a 
non perturbative level due to the Gribov ambiguity \cite{Gribov}  a
unique 
gauge fixing in the full space of gauge fields is not possible so that
the 
Faddeev-Popov path integral procedure is  defined only locally in field 
space. 

Several attempts  were presented to generalize   the above approach 
in order to establish global integral 
representations \cite{Stora,Becchi} in the space of gauge fields. 
It is our aim to present a quite different 
argumentation based on a recently introduced
generalized stochastic quantization scheme
\cite{Physlett,annals,annalsym,global}. 

The stochastic quantization method of Parisi and Wu  \cite{Parisi+Wu}
was introduced 
1981 as a new method for quantizing field theories. It is based on 
concepts of nonequilibrium statistical mechanics and provides novel and
alternative 
insights into quantum field theory, see refs.
\cite{Damgaard+Huffel,Namiki}.
 One of the most interesting aspects of this new quantization 
scheme lies in 
its rather unconventional treatment of gauge field theories, in specific
of
Yang-Mills theories. We  just recall that 
originally it was 
formulated  by Parisi and Wu  without 
the introduction of gauge fixing terms and the usual 
Faddeev-Popov ghost 
fields; later on a modified approach named stochastic gauge fixing
was given by Zwanziger \cite{Zwanziger81} where again no Faddeev-Popov 
ghost fields where introduced. 
Our focus is based 
on extending  Zwanziger's 
stochastic gauge fixing 
scheme.  By this 
generalized stochastic gauge fixing scheme it is possible to derive 
a non perturbative proof of the equivalence between  
the conventional path integral formulation of this model and the 
equilibrium limit of the corresponding stochastic correlation functions. 

The main difficulty in the previous investigations of the stochastic 
quantization of Yang--Mills theory for  deriving a conventional field
theory 
path integral density was to solve the Fokker--Planck equation 
in the equilibrium limit. 
In the original Parisi--Wu approach this equilibrium limit could not
even be 
attained due to unbounded diffusions of the gauge modes.
Zwanziger \cite{Zwanziger81,Baulieu+Zwanziger}
 suggested to introduce a specific additional  nonholonomic
stochastic force term to 
suppress these gauge modes 
yet keeping the expectation values of 
gauge invariant observables unchanged. The approach to equilibrium and 
the discussion of the 
conditions of applicability to the nonperturbative regime, 
however, do not seem to have been fully completed.

Our analysis is distinguished by the above approaches 
by exploiting a more general freedom to modify 
both the drift term and the diffusion term of the stochastic process
again leaving all expectation values 
of gauge invariant variables unchanged. Due to this additional structure 
of modification the equilibrium limit can be obtained immediately using 
the fluctuation dissipation theorem proving equivalence with the well 
known Faddeev-Popov path integral density. In deriving this result the 
gauge degrees of freedom are fully under control, no infinite gauge 
group volumes arise. 
However, this equivalence proof can be performed 
only locally in field space for  gauge field configurations 
satisfying a unique gauge 
fixing condition and deserves an extension for global applicability. We
are able 
to discuss such a formulation in the second half of this paper.

Let $P(M,G)$ be a principal fiber 
bundle with   compact structure group $G$ over
the compact Euclidean space time $M$. Let $\A$ denote the space of all
irreducible connections  on $P$ and let $\G$ denote the gauge group, 
which is given by all vertical automorphisms on $P$ reduced by the
centre
of $G$. Then $\G$ acts freely on $\A$ and defines a principal 
$\G$-fibration
$\A \st{\pi}{\longrightarrow} \A/\G =: \M$ over the paracompact 
\cite{Mitter} space
$\M$ of all inequivalent gauge potentials
with projection $\pi$. 
Due to the Gribov ambiguity the principal 
$\G$-bundle $\A \ra \M$
is not globally trivializable.

Using this mathematical setting   we
start with the Parisi--Wu approach for the stochastic quantization
of the Yang--Mills theory in terms of the Langevin equation
\beq
dA = - \frac{\delta S}{\delta A} ds + dW.
\eeq
Here $S$ denotes the Yang--Mills action without gauge symmetry breaking
terms and without accompanying ghost field terms,
$s$ denotes the extra time coordinate (``stochastic time'' coordinate)
with respect to which the stochastic process is evolving, $dW$ is the
increment of a Wiener process.

We now discuss Zwanziger's modified formulation  of 
the Parisi-Wu scheme: The stochastic gauge fixing procedure 
consists in adding an additional drift force to the Langevin 
equation (1) which acts tangentially to the gauge orbits. This 
additional term generally can be expressed by the gauge 
generator $D_A$, i.e. the covariant derivative with respect to $A$, 
 and an arbitrary function $\bf a$
 so that the modified Langevin equation reads as follows
 \beq
dA = \left[ -\frac{\delta S}{\delta A} + D_A {\bf a} 
\right] ds
+ dW.
\eeq
The expectation values of gauge invariant observables remain 
unchanged for any choice of the function $\bf a$ .
For specific choices of the -- in principle -- arbitrary 
function 
$\bf a$ the gauge modes' diffusion is damped along the gauge 
orbits. As a consequence the Fokker-Planck density can be 
normalized; we remind that this situation is different to the Parisi-
Wu approach, where for expectation values of gauge variant 
observables no equilibrium values could be attained.

In contrast to the approach of \cite{Zwanziger81} where no equilibrium 
distribution of the Fokker--Planck equation could be derived as well as 
in contrast to \cite{Baulieu+Zwanziger} where the full Fokker--Planck 
operator 
\beq
L = \frac{\delta }{\delta A}\left[  \frac{\delta S}{\delta A} - D_A 
{\bf a} + \frac{\delta }{\delta A} 
\right] 
\eeq       
was needed to obtain an equilibrium distribution we present a 
quite different strategy:

As the Fokker-Planck operator factorizes into first order differential 
operators  the question arises whether it is possible to derive the 
equilibrium distribution directly by solving a simpler first order 
problem. However, for this to be possible a necessary integrability 
condition imposed on the drift term $\dfrac{\delta S}{\delta A} - D_A 
\bf a$ has to be fulfilled. It is 
well known that for the Yang--Mills case this is violated.   

In the following we want to clarify the relationship of this 
integrability condition and 
the underlying geometrical structure of the space of gauge potentials.
We remind that any bundle metric on a principal fiber
bundle which is invariant under the corresponding group action gives
rise
to a natural connection whose horizontal subbundle is orthogonal to the
corresponding group. In the Yang-Mills theory case with respect to 
the natural metric on $\A$ this  
connection  is given by the 
following $Lie\G$ valued one form
\beq
\gamma = \Delta_A^{-1} D_A^*.
\eeq
 Here 
$D_A^*$ 
is the
adjoint operator of the covariant derivative $D_A$,  $\Delta_A^{-1}$
is the inverse of the covariant Laplacian 
$\Delta_A = D_A^* D_A$.

The curvature $\Omega$ of $\gamma$, $\Omega = \delta_\A \gamma + 
\frac{1}{2} [\gamma,\gamma]$, where $\delta_\A$ denotes the exterior 
derivative on $\A$, however, does not vanish  so that there does
not exist (even locally) a manifold whose tangent bundle is isomorphic
to this horizontal subbundle. Moreover this also implies that any vector
field along the gauge group cannot be written as a gradient of a
function.

It is our intention to modify the stochastic process (1) for the 
Yang--Mills theory in such a 
way that the factorization of the modified Fokker-Planck operator indeed
allows 
the determination of the equilibrium distribution as a solution of a
first 
order differential equation in a consistent manner.

From \cite{Mitter} it follows that there exists a locally finite open 
cover $\U
=\lbrace U_{\alpha} \rbrace$  of $\M$ together with a set of background 
gauge fields $\lbrace A_{0}^{(\alpha) } \in \A
 \rbrace$  such that 
\beq
\Gamma_{\alpha} = \{ B \in 
\pi^{-1} (U_{\alpha})|D^{*}_{A_{0}^{(\alpha)}} (B - 
A_{0}^{(\alpha)}) = 0\}
\eeq
defines a family of local sections of $\A \ra \M$.
Instead of analyzing Yang-Mills theory in the original field 
space $\A$  we 
consider the familiy of trivial  principal $\G$-bundles 
$\Gamma_{\alpha} \times \G \ra \Gamma_{\alpha}$, 
which are locally isomorphic to the bundle 
$\A \ra \M$, where the isomorphisms are provided by the maps
\beq
\chi_{\alpha} : \Gamma _{\alpha} \times \G \ra 
\pi^{-1}(U_{\alpha}), \qquad
\chi_{\alpha} (B,g) := B^g
\eeq
with $B \in \Gamma_{\alpha}$, $g \in \G$ and 
$B^g$ denoting the  gauge 
transformation of $B$ by $g$. 
Thus we transform the Parisi--Wu Langevin equation (1) into 
the adapted coordinates $\Psi = \left( \ba{c} B \\ g \ea \right)$. As 
this transformation is not globally possible the region of 
definition of (1) has to be restricted to
$\pi^{-1}(U(A_{0}^{(\alpha)}))$. Making use of the Ito stochastic
calculus
 the above Langevin equation now reads 
\beq
d\Psi = \left( - G_{\alpha}^{-1} \frac{\delta S_{\alpha}}{\delta \Psi} + 
\frac{1}{\sqrt{\det G_{\alpha}}} \frac{\delta(G_{\alpha}^{-1}
 \sqrt{\det G_{\alpha}})}{\delta \Psi}
\right)ds + E_{\alpha} dW
\eeq
where $S_{\alpha}=\chi_{\alpha}^* S$ denotes the gauge invariant 
Yang-Mills action expressed in terms of the adapted coordinate $B$, and 
where the explicit forms of the vielbein $E_{\alpha}$ corresponding to
the change 
of coordinates $A \ra (B,g)$, the induced metric $G_{\alpha}$, 
its inverse and its determinant
can be found in \cite{annalsym}; for completeness we just recall that
\beq
\det G_{\alpha} = \det (R_g^* R_g) \, (\det \F_{\alpha})^2 \,
(\det \Delta_{A_{0}^{(\alpha)}})^{-1}.
\eeq
Here $\sqrt{\det (R_g^* R_g)}$ implies an invariant volume density 
on $\G$, 
where $R_g$  is the
differential of right multiplication
transporting any tangent vector in $T_g \G$ back to the identity 
$id_{\G}$  on $\G$ ; 
$\F_{\alpha} = D_{A_{0}^{(\alpha)}}^* D_B$
is the Faddeev--Popov operator.

The generalized stochastic quantization procedure amounts
 to consider the modified Langevin equation
\beq
d\Psi = \left( -G_{\alpha}^{-1} \frac{\delta S_{\alpha}}{\delta \Psi} + 
\frac{1}{\sqrt{\det G_{\alpha}}} \frac{\delta(G_{\alpha}
^{-1}\sqrt{\det G_{\alpha}})}{\delta\Psi}
 + E_{\alpha} D_A {\bf a} \right) ds + E_{\alpha}({\bf 1} + D_A {\bf
b})dW,  
\eeq
where $A=B^g$. Here $\bf a$ 
and the $Lie \G$ valued one form 
$\bf b$
are \`a priori arbitrary and will be fixed later on. 

The above Langevin 
equation is the most general 
Langevin equation for Yang--Mills theory which leads to the same 
expectation values of gauge invariant variables as the original 
Parisi--Wu equation (1):
An easy way to prove this assertion is to  observe that  the $\bf a$ and
$\bf b$ dependent terms in the modified Langevin equation (9) drop
out after projecting on the gauge invariant subspace $\Gamma_{\alpha}$
described 
by the coordinate $B$.

Transforming back the Langevin equation (9) into the
original coordinates $A$
not only Zwan\-ziger's original term $D_A \bf a$ is appearing, but also
an
additional $\bf b$-dependent drift term as well as a specific
modification
of the Wiener increment, described by the operator $\wh e = {\bf 1} + 
D_A {\bf b}$. The idea is to view
$\wh e$ as a vielbein giving rise to the inverse of a  metric $\wh g$ on 
the space $\pi^{-1}(U(A_{0}^{(\alpha)}))$.
Since any of the $\wh g$ (parametrized by the not yet specified 
$\bf b$) 
implies a 
specific connection one is likely to arrive at an analogous obstruction
as 
mentioned
above.  
It is therefore necessary to require
that the corresponding curvature should vanish.
Indeed, we know  that locally there exists a flat connection $\wt 
\gamma_{\alpha}$ in our 
bundle namely the pull-back of the Maurer--Cartan form
$\theta = \mbox{ad}(g^{-1})R_g$ on the gauge group via the 
local trivialization 
of the bundle $\pi^{-1}(U(A_{0}^{(\alpha)})) 
\ra U(A_{0}^{(\alpha)})$. 
Explicitely   $\wt \gamma_{\alpha}$ is given by the following expression
\beq
\wt \gamma _{\alpha}
 = \mbox{ad}(g^{-1}) \F_{\alpha}^{-1} D_{A_{0}^{(\alpha)}}^*
\mbox{ad}(g), \qquad
A = B^g.
\eeq 
The associated horizontal subbundle $\wt \Ha$ is built by all 
those vectors  in the tangent space in 
$A \in \pi^{-1}(U(A_{0}^{(\alpha)})) $ 
 which can be written in the form $\tau = \mbox{ad}(g^{-1})
\zeta_B$, where $A = B^g$ and $\zeta_B$  is a tangent vector
of $\Gamma_{\alpha}$ in point $B$.
The fact that the curvature corresponding to the connection
$\wt \gamma _{\alpha}$ is vanishing implies that 
 the horizontal subbundle $\wt \Ha$ is
 integrable. The connection $\wt \gamma_{\alpha}$ cannot be 
extended to a globally defined flat connection on the whole bundle $\A 
\ra \M$, however, 
due to its 
nontriviality. 
 
We  determine the value of $\bf b$ by fixing the 
 metric $\wh g$ in such a way that  $\wt \gamma _{\alpha}$ 
 is exactly the induced connection
imposed by itself. This implies that the {\it horizontal
subbundle $\wt \Ha$ is orthogonal to the gauge orbits
with respect to the metric $\wh g$} and 
-in this sense- {\it the gauge fixing surface is 
orthogonal to the gauge orbits.}

The determination of $\bf a$ can in fact be given a suggestive meaning,
too:  
we have the freedom 
to totally exchange the drift term of 
the $g$-field component of the Langevin equation  (9)
 by the damping term $- (\wt G_{\alpha}^{-1})^{\G\G}\dfrac{\delta 
S_{\G}[g]}{\delta g}$ and  add a judiciously chosen Ito--term 
$\dfrac{1}{\sqrt{\det G}} 
\dfrac{\delta((\wt G^{-1})^{\G\G}\sqrt{\det G})}{\delta g}$ as well.
Here we introduced a new vielbein $\wt E_{\alpha}$ and a new (inverse)
metric 
$\wt G_{\alpha}^{-1}$
\beq
\wt E_{\alpha} = E_{\alpha}({\bf 1} + D_A {\bf b}), \qquad 
\wt G_{\alpha}^{-1} = \wt E_{\alpha} \wt E_{\alpha}^*, 
\eeq
with $A=B^g$; furthermore $S_{\G}[g]$ is an arbitrary damping function
with the
property that
\beq
\int_{\bf \G} \D g \sqrt{\det(R_g^* R_g)} \;
e^{-S_{\G}[g]} < \infty.
\eeq
Due to these choices for $\bf a$  and $\bf b$ we firstly   obtain a
well damped Langevin equation for $g$ and secondly can recast 
the Langevin equation (9)  
 into the geometrically distinguished form
\beq
d\Psi = \left[- \wt G_{\alpha}^{-1} \frac{\delta S_{\alpha}^{\rm tot}
}{\delta \Psi}
+ \frac{1}{\sqrt{\det G_{\alpha}}} 
\frac{\delta(\wt G_{\alpha}^{-1} \sqrt{\det G_{\alpha}
})}{\delta \Psi} \right] ds
+ \wt E_{\alpha} dW.
\eeq
Here
\beq
S_{\alpha}^{\rm tot} = \chi_{\alpha}^* S + pr_{\G}^* S_{\G}
\eeq
denotes a total Yang-Mills action  
defined by the original Yang-Mills action $S$
without gauge symmetry breaking terms and by the above $S_{\G}$;  
$pr_{\G}$ is the projector $\Gamma_{\alpha} \times \G \ra \G$.
The associated Fokker--Planck equation  can be derived in a 
straightforward manner
where now the Fokker-Planck operator $L[\Psi]$ is 
appearing in the factorized form
\beq
L[\Psi] = \frac{\delta}{\delta \Psi}
\wt G_{\alpha}^{-1} \left[ \frac{\delta S_{\alpha}^{\rm tot}}{\delta
\Psi}
- \frac{1}{\sqrt{\det G_{\alpha}}} 
\frac{\delta(\sqrt{\det G_{\alpha}})}{\delta \Psi}
+ \frac{\delta}{\delta \Psi}
 \right].
\eeq
Due to the positivity of $\wt G_{\alpha}$ the fluctuation dissipation
theorem 
applies and the (non-normalized) equilibrium Fokker--Planck distribution 
can be 
obtained by direct inspection of the first order differential 
operator on the right hand side of the Fokker-Planck operator as
\beq
\mu_{\alpha} \, e^{-S_{\alpha}^{\rm tot}}, \quad 
\mu_{\alpha} = \sqrt{\det G_{\alpha}}.
\eeq
It is the basic idea of the  stochastic quantization scheme to interpret
an equilibrium limit of 
a Fokker--Planck distribution as  Euclidean path integral measure. 
Although our above result for the path integral measure implies 
unconventional 
$finite$ contributions along the gauge group (arising from the 
$pr_{\G}^* S_{\G}$ term) it is equivalent to the usual 
Faddeev--Popov prescription for Yang--Mills theory. This follows from 
the fact that for expectation 
values of gauge invariant observables these contributions along the
gauge 
group are exactly cancelled out due to the normalization of the path
integral, see 
below. We stress once more that 
due to the Gribov ambiguity the usual Faddeev--Popov 
approach as well as -presently- our modified version 
are valid only locally in field space. 

In order to compare  expectation values on different patches we 
consider the  diffeomorphism in the overlap of two patches 
\beq
\phi_{\alpha_1 \alpha_2} : (\Gamma_{\alpha_1} \cap
\pi^{-1}(U_{\alpha_2})) \times \G 
\ra 
(\Gamma_{\alpha_2} \cap \pi^{-1}(U_{\alpha_1})) \times \G
 \qquad
\phi_{\alpha_1 \alpha_2} (B,g) := 
(B^{\omega_{\alpha_2}(B)^{-1}}, g).
\eeq
Here $\omega_{\alpha_2} : \pi ^{-1}(U_{\alpha_2}) \ra \G$ is 
uniquely defined by 
$A^{\omega_{\alpha_2}(A)^{-1}} \in \Gamma_{\alpha_2}$. 
To the  density  $\mu_{\alpha}$ there is associated  a corresponding 
twisted top  form on $\Gamma_{\alpha} \times \G$ 
(see e.g. \cite{Bott}) which for simplicity we denote by the same
symbol.
Using for convenience a matrix 
representation of $G_{\alpha}$ \cite{annalsym} we 
straightforwardly verify that
\beq
\phi_{\alpha_1 \alpha_2}^* \, \mu_{\alpha_2} = \mu_{\alpha_1} \, .
\eeq
This immediately implies  that in overlap regions the 
 expectation values of gauge invariant observables $f \in
C^{\infty}(\A)$ 
 are equal 
when evaluated in different  patches

Finally we propose the definition of the global expectation value of a 
gauge 
invariant observable $f \in C^{\infty}(\A)$ by summing over all the 
elements $e_{\alpha}$ of a
partition of unity on $\M$ (the existence of  a partition of 
unity is guaranteed by the result of \cite{Mitter})
so that  
\beq
\langle f \rangle = 
\frac{\sum_{\alpha} \lint_{\Gamma_{\alpha} 
\times \G} \mu_{\alpha} \, e^{-S_{\alpha}^{\rm tot}}
\chi_{\alpha}^* (f \pi ^* e_{\alpha})}
{\sum_{\alpha} \lint_{\Gamma_{\alpha} \times \G} 
\mu_{\alpha} \, e^{-S_{\alpha}^{\rm tot}}
\chi_{\alpha}^*  \pi ^* e_{\alpha} }.
\eeq
Due to (18) it is trivial to prove that the global 
expectation value $\langle f 
\rangle$ is independent of  the specific 
choice of the locally finite cover $\lbrace U_{\alpha} \rbrace$, 
of the  choice of the background 
gauge fields $\lbrace A_{0}^{(\alpha)} \rbrace$ 
and of the choice of the partition of unity $e_{\alpha}$, 
respectively.

These structures can equally be 
translated into the original field space $\A$. With the help 
of the partion of unity the locally 
defined densities $\mu_{\alpha}$ as well as $e^{-S_{\alpha}^{\rm tot}}$
can be pieced together to give a globally well defined 
twisted top form $\Omega$ on $\A$
\beq
\Omega := \sum_{\alpha}  \chi_{\alpha}^{-1 \, *}
(\mu_{\alpha} \,  e^{-S_{\alpha}^{\rm tot}}) \pi ^* e_{\alpha}. 
\eeq
The  global expectation value (19) then reads
\beq
\langle f \rangle = 
\frac{ \int_{\A} \Omega \, f} 
{\int_{\A} \Omega }
\eeq
which due to the discussion from above is independent of all
the particular local choices.

Let us extract the gauge invariant part of the local Fokker--Planck
densities 
(16)
\beq
\det \F_{\alpha} \,
(\det \Delta_{A_{0}^{(\alpha)}})^{-1/2} \,
e^{-\chi_{\alpha}^* S}.
\eeq
By using (18) we can prove that their projections on $\M$ on overlapping 
sets of $\U$ are agreeing so that they are
 giving rise to a globally well defined top 
form  $\tilde{\Omega}$ on 
$\M$. One can furthermore show that the above expectation values  
of  gauge invariant 
functions $f$ can identically be rewritten as  corresponding integrals
over 
the gauge orbit space $\M$ with respect to $\tilde{\Omega}$
\beq
\langle f \rangle = 
\frac{ \int_{\M} \tilde{\Omega} \, f} 
{\int_{\M} \tilde{\Omega} }.
\eeq
We note that this last expression shows agreement with the formulation 
proposed by Stora \cite{Stora} upon identification of $\tilde{\Omega}$
with the Ruelle-Sullivan form  \cite{Ruelle}. 

Whereas in \cite{Stora} the  global definition (23) of 
expectation values on 
$\M$ appeared as the starting point 
for a path integral formulation of Yang-Mills theory in the whole space
of 
gauge potentials it appears now as our final result; 
we  aimed 
at its direct
derivation  within the stochastic quantization approach: 
First we derived a  local path integral measure on $\Gamma_{\alpha}
\times \G$
 in terms of the probability density 
$\mu_{\alpha} \, e^{-S_{\alpha}^{\rm tot}}$ which 
 assured   gaussian 
decrease along the gauge fixing surface
$as \, well \, as$ along the gauge orbits. The inherent 
interrelation 
of the field variables on the  patches $\Gamma_{\alpha} \times \G$  
subsequently  
led to  simple relations of the local
 densities  in the overlap regions and eventually
to  the global  path integral formulations (19), (21) and (23),
respectively.
 
H. H{\"u}ffel acknowledges generous support from the organizers of the 
Central European Triangle 
Meeting on Particle Physics during his stay in Zagreb; G. Kelnhofer
acknowledges 
support by "Fonds zur F\"orderung der wissenschaftlichen 
Forschung in \"Oster\-reich",  project P10509-NAW.

\end{document}